\documentclass[12pt]{iopart} 
\usepackage{epsfig}
\usepackage{graphicx}

\newcommand{\be}{\begin{equation}} 
\newcommand{\ee}{\end{equation}} 
\newcommand{\bd}{\begin{displaymath}} 
\newcommand{\ed}{\end{displaymath}} 
\newcommand{\vsp}{\vspace*{3mm}}

\newcommand{\bra}{\langle} 
\newcommand{\ket}{\rangle}

\newcommand{\order}{{\cal O}}

\newcommand{\bsigma}{{\mbox{\boldmath $\sigma$}}}

\newcommand{\bc}{{\bf c}}

\newcommand{\bxi}{{\mbox{\boldmath $\xi$}}}

\newcommand{\cP}{{\mathcal P}}

\newcommand{\odis}{\big<\hspace{-1mm}\big<} 
\newcommand{\cdis}{\big>\hspace{-1mm}\big>} 
\newcommand{\Odis}{\Big<\hspace{-1.5mm}\Big<} 
\newcommand{\Cdis}{\Big>\hspace{-1.5mm}\Big>}

\begin{document} 
 
\title{Analytic solution of attractor neural networks on scale-free graphs} 
 
\author{I P\'{e}rez Castillo$^\ddag$, B Wemmenhove$^\P$, J P L Hatchett$^\dag$,  A C C Coolen$^\dag$,  N S Skantzos$^\S$, 
  and T Nikoletopoulos$^\dag$} 
\address{\ddag ~ Institute for Theoretical Physics, Celestijnenlaan 200D, Katholieke Universiteit Leuven, B-3001 Belgium} 
\address{\P ~Institute for Theoretical Physics, University of Amsterdam, Valckenierstraat 65,1018 XE Amsterdam, The Netherlands} 
\address{\dag ~ Department of Mathematics, King's College London, The Strand,London WC2R 2LS, United Kingdom} 
\address{\S ~ Departament de F\'\i sica Fonamental, Facultat de F\'\i sica, Universitat de Barcelona, 08028 Barcelona, Spain} 
 
\begin{abstract} 
We study the influence of network topology on retrieval 
properties of recurrent neural networks, using replica techniques for 
diluted systems. The theory is presented for a network with an 
arbitrary degree distribution $p(k)$ and applied to power law distributions 
$p(k) \sim k^{-\gamma}$, i.e. to neural networks on scale-free graphs. 
A bifurcation analysis 
identifies phase boundaries between 
the paramagnetic phase and either a retrieval phase or a spin glass 
phase. Using a population dynamics algorithm, the retrieval overlap 
and spin glass order parameters may be calculated throughout 
the phase diagram. It is shown that there is an 
enhancement of the retrieval properties compared with a Poissonian 
random graph. We compare our findings with simulations. 
\end{abstract} 
 
 \ead{\tt 
isaac.perez@fys.kuleuven.ac.be,
wemmenho@science.uva.nl, hatchett@mth.kcl.ac.uk, 
 tcoolen@mth.kcl.ac.uk, nikos@ffn.ub.es, theodore@mth.kcl.ac.uk} 
 
\section{Introduction} 
 
The impressive ability of human and animal brains to recognize and 
manipulate complex patterns under real-world (i.e. noisy and often 
conflicting) conditions continues to appeal not only to biologists 
but also to physicists, computer scientists and engineers, albeit 
with the latter 
 driven by different objectives and motivations. 
Hopfield \cite{Ho82} was one of the first to introduce a simple 
model to describe associative memory in recurrent neural networks 
successfully, based on the biologically motivated Hebbian rule for 
adapting the connections between the neurons (the `synapses'). His 
model initiated 
 a period of intense research activity. 
 The success of these early neural network models 
was mainly due to their analytic tractability, which was achieved 
upon sacrificing  biological realism; all neurological 
connectivity structures were sacrificed by the first generation of 
(fully connected) models. However, the research area was thereby 
able to  benefit significantly from recent advanced in mean-field 
spin glass theory \cite{MPV,codesrefNishimori}, and many new 
results were published in the second half of the eighties; see 
e.g. \cite{AGS1,AGS2,GAR} or \cite{DHS1}. 
 
A step towards increased biological realism  was made by the 
introduction of diluted neural network models. Initially, in the 
thermodynamic limit each neuron was on average connected to a 
vanishing fraction of the system, but this fraction contained an 
infinite number of nodes. These models were solvable by virtue of 
the specific nature of their architectures: one either chooses 
strictly symmetric dilution (so detailed balance and hence 
equilibrium analysis are preserved, e.g. \cite{WS1,WS2,CaNe}), or 
strictly asymmetric dilution, which ensures that neuron states are 
statistically independent on finite times \cite{DGZ} (now the 
local fields are described by Gaussian distributions, leading to 
simple dynamic order parameter equations). In the early models, 
the bond statistics were uniform over the entire network, leading 
to thin tails in its degree distribution, whereas the connectivity 
of a real neuron is known to vary strongly within the brain 
\cite{EgChCeBaAp03}. In response to this, there have been several 
recent studies of recurrent neural network models with alternative 
connectivity distributions. Most evolve around numerical 
simulations of Hopfield-type models on graphs with power law 
degree distributions 
\cite{ToMuMaGa03,ToMaGaCoRaMu03,McMe03,StAhFoAd02}. Examples of 
recent analytic work on recurrent neural networks with finite 
connectivity can be found in \cite{WeCo03,PeSk03}; both deal with 
Poissonian graphs and apply the equilibrium statistical mechanical 
techniques of dilute disordered spin systems 
\cite{VB85,KS87,MP87}. 
 
The objective of this paper is to extend and generalize the 
solution for finitely connected Poissonnian neural networks 
\cite{WeCo03,PeSk03} to recurrent neural networks with arbitrary 
degree distribution, in the spirit of \cite{WoSh87,LeVaVeZe02} and 
within the replica-symmetric (RS) ansatz. We derive analytically 
phase diagrams for networks with Hebbian synapses and arbitrary 
degree distributions $p(k)$, and apply population dynamics 
algorithms to obtain the values of the order parameters in the 
three phases (viz. paramagnetic, retrieval, and spin-glass). 
 This study thereby 
establishes a connection between the equilibrium statistical 
mechanics of neural networks and the theory of so-called `complex 
networks'. In line with biological reality, we find that recurrent 
neural networks with degree distributions with `fat tails' are 
indeed superior to those with Poissonnian degree distributions, in 
terms of the size of the recall region in the phase 
diagram. 
 
\section{Model definitions} 
\label{definitions} 
 
Our model is a system of $N$ Ising spin neurons 
 $\sigma_i\in\{-1,1\}$, with $i=1,\ldots,N$. The neurons are located 
 on the nodes of a graph with arbitrary degree distribution $p(k)=N^{-1}\sum_i\delta_{k,k_i}$, where $k_i$ 
 denotes the number of neurons connected to neuron $i$. This system is assumed to be in thermodynamic equilibrium, described 
 by the Hamiltonian 
\begin{equation} 
{\cal H}=-\sum_{i<j=1}^N\sigma_iJ_{ij}\sigma_j-\sum_{\mu=1}^p 
h^\mu\sum_{i=1}^N\xi^\mu_i\sigma_i \label{eq:Hamiltonian} 
\end{equation} 
Here the $\{h^\mu\}$ represent generating fields, and the vectors 
$(\xi_1^\mu,\ldots,\xi_N^\mu)$ represent stored random $N$-bit 
patterns. We will abbreviate the bits to be stored at a given node 
$i$ as  $\bxi_i=(\xi_i^1,\ldots,\xi_i^p)$. Since the number of 
connections per neuron is finite, the number of patterns $p$ must 
be of order $\mathcal{O}(N^0)$. The bonds $J_{ij}$ depend on the 
patterns via 
\begin{equation} 
J_{ij}=\frac{c_{ij}}{\langle k \rangle}\phi\big(\bxi_i\cdot\bxi_j\big) 
\end{equation} 
with  $\bxi_i\cdot\bxi_j=\sum_{\mu=1}^p\xi^\mu_i\xi^\mu_j$ and 
$\langle k \rangle = \sum_{k\geq 0} p(k) k$. Special cases of 
interest are $\phi(x)=x$, viz. Hebbian bonds, and $\phi(x) = 
\mbox{sign}(x)$, viz. clipped Hebbian bonds. 
 
The variables $c_{ij} \in \{0, 1\}$ specify our graph 
microscopically; we extend their definition to all pairs $(i,j)$ 
by putting $c_{ij}=c_{ji}$ and $c_{ii}=0$. It is known from 
complex network theory \cite{GrSnMi04,FarDer} that a connectivity 
distribution $p(k)$ alone does not fully specify the statistics of 
a graph. Here we draw the matrix $\bc=\{c_{ij}\}$, which 
represents quenched disorder for the spin system 
(\ref{eq:Hamiltonian}), randomly from the probability distribution 
\be 
\cP(\bc)=\frac{\Big[\prod_{i<j}P(c_{ij})~\delta_{c_{ij},c_{ji}}\Big]\Big[\prod_{i}\delta_{k_i,\sum_{j\neq 
i} c_{ij}}\Big]} 
{\sum_{\bc^\prime}\Big[\prod_{i<j}P(c^\prime_{ij})~\delta_{c^\prime_{ij},c^\prime_{ji}}\Big]\Big[\prod_{i} 
\delta_{k_i,\sum_{j\neq i} c^\prime_{ij}}\Big]} 
\label{eq:ensemble} 
 \ee with the 
single-bond probabilities 
\begin{eqnarray} 
P(c_{ij})& = & \frac{\langle k 
\rangle}{N}\delta_{c_{ij},1}+\Big(1-\frac{\langle k 
\rangle}{N}\Big)\delta_{c_{ij},0} 
\end{eqnarray} 
Disorder averages $\odis A(\bc)\cdis_{\bc}$ over the ensemble of 
graphs are thus given by \cite{WoSh87,LeVaVeZe02} 
\begin{eqnarray} 
\odis A(\bc)\cdis_{\bc}&=&{\cal N}^{-1} \sum_{\bc} 
\Big[\prod_{i<j} P(c_{ij})~\delta_{c_{ij},c_{ji}}\Big] \Big[ 
\prod_{i}\int\!\frac{d\psi_i}{2\pi}~e^{i\psi_i( \sum_j 
c_{ij}-k_i)}\Big] 
 A({\bf c}) 
 \label{eq:dis_av} 
\\ 
\qquad {\cal N} &=&\sum_{\bc} \Big[\prod_{i<j} 
P(c_{ij})~\delta_{c_{ij},c_{ji}}\Big] \Big[\prod_{i} 
\int\!\frac{d\psi_i}{2\pi}~e^{i\psi_i( \sum_j c_{ij}-k_i)}\Big] 
\end{eqnarray} 
 Another statistical quantity to  characterize  graphs, beyond $p(k)$, is the degree-degree correlation: the 
joint probability $\omega(k_i,k_j)$ that a pair of nodes $i$ and 
$j$ are connected, and  have  connectivities $k_i$ and $k_j$, 
respectively. For the present ensemble (\ref{eq:ensemble}) one 
finds 
\begin{equation} 
\omega(k_i, k_j) = \frac{p(k_i) p(k_j) k_i k_j}{\langle k \rangle 
N} \label{eq:degdegcor} 
\end{equation} 
 
\section{Replica calculation of the free energy and order parameters} 
\label{replica} 
 
We calculate the free energy per spin and the relevant order 
parameters, using the replica techniques as developed for 
constrained connectivity graphs,   along the lines of 
\cite{LeVaVeZe02,WoSh87}. Thus the asymptotic free energy per spin 
$f=-\lim_{N\to\infty}(\beta N)^{-1}\log {\cal Z}$ is written as 
\begin{equation} 
 f=-\lim_{N\to\infty}\lim_{n\to0}\frac{1}{Nn \beta}\log\odis{\cal Z}^n\cdis_{{\bf c}} 
\end{equation} 
where ${\cal Z}^n$ is the usual $n$-replicated partition function 
\begin{equation} 
{\cal 
Z}^n=\sum_{\bsigma_1}\cdots\sum_{\bsigma_N}\exp\Big[\beta\sum_{\alpha=1}^n\sum_{i<j}\sigma_i^\alpha 
J_{ij}\sigma_j^\alpha +\beta\sum_{\mu=1}^p 
h^\mu\sum_{i=1}^N\sum_{\alpha=1}^n\xi^\mu_i\sigma_i^\alpha\Big] 
\label{eq:Zn} 
\end{equation} 
and $\bsigma_i\equiv(\sigma_i^1,\ldots,\sigma_i^n)$ is the 
$n$-replicated spin at site $i$. Upon performing the trace over 
the $c_{ij}$ (i.e. the disorder average) one obtains 
\begin{eqnarray} 
 \odis{\cal Z}^n\cdis_{{\bf c}} &=& \frac{1}{{\cal 
N}}\prod_{i=1}^N\left[\sum_{\{\bsigma_i\}}\int 
\frac{d\psi_i}{2\pi}e^{-i\psi_ik_i}\right]\exp\Big[\beta\sum_{\mu=1}^p 
h^\mu\sum_{i=1}^N\sum_{\alpha=1}^n\xi_i^\mu\sigma_i^\alpha\Big]\nonumber\\ 
&&\times~\exp\Big[\frac{\langle 
k\rangle}{2N}\sum_{i,j=1}^N\Big(e^{i(\psi_i+\psi_j)+\frac{\beta}{\langle 
k\rangle}\phi(\bxi_i\cdot\bxi_j)\bsigma_i\cdot\bsigma_j}-1\Big)+\order(N^0)\Big] 
\end{eqnarray} 
In order to go to an effective single-site problem one first 
introduces the concept of sublattices $I_{\bxi}=\{i|\bxi_i=\bxi\}$ 
and defines the following order parameter functions: 
\begin{equation} 
P_{\bxi}(\bsigma)=\frac{1}{|I_{\bxi}|}\sum_{i\in 
I_{\bxi}}e^{i\psi_i}\delta_{\bsigma,\bsigma_i} \label{eq:P_unnorm} 
\end{equation} 
These are reminiscent of the replicated spin probability 
distributions within sublattices, as in \cite{WeCo03}, but here 
include extra phase factors $e^{i\psi_i}$ whose effect is to 
replace $k_i$ by $k_i-1$ in expressions of the type 
(\ref{eq:dis_av}). At the physical saddle-point in the subsequent 
calculation one finds the physical meaning of (\ref{eq:P_unnorm}) 
to be 
\begin{equation} 
P_{\bxi}(\bsigma)=\frac{1}{|I_{\bxi}|}\sum_{i\in I_{\bxi}} \Odis 
\bra  \delta_{\bsigma,\bsigma_i}e^{i\psi_i}\ket \Cdis_{\bc} 
\end{equation} 
with $\bra \ldots\ket$ denoting a thermal average over the 
$n$-replicated spins.  Thus the order parameter 
(\ref{eq:P_unnorm}) is just the distribution of a replicated {\em 
cavity} spin in sublattice $\bxi$. After some straightforward 
manipulations (factorization over sites, integration over the 
variables $\psi_i$, etc) we find $f = \lim_{n \to 0} {\rm 
extr}_{\{P, \hat{P}\}} f[\{ P, \hat{P}\}]$, where 
\begin{eqnarray} 
\hspace*{-10mm} f[\{ P, \hat{P}\}] & = & \frac{\langle 
k\rangle}{\beta n} 
\sum_{\bsigma}\bra\widehat{P}_{\bxi}(\bsigma)P_{\bxi}(\bsigma)\ket_{\bxi} 
-\frac{1}{\beta n}\sum_{k}p_k\bra\log\Big[\sum_{\bsigma}e^{\beta 
{\bf h}\cdot\bxi 
\sum_{\alpha=1}^n\sigma_\alpha}\widehat{P}^{k}_{\bxi}(\bsigma)\Big]\ket_{\bxi} 
\nonumber 
\\ 
 &&-\frac{\langle k\rangle }{2\beta n}\sum_{\bsigma,\bsigma^\prime}\bra\bra 
P_{\bxi}(\bsigma)P_{\bxi^\prime}(\bsigma^\prime)e^{\frac{\beta}{\langle 
k\rangle}\phi(\bxi\cdot\bxi^\prime)(\bsigma\cdot\bsigma^\prime)}\ket\ket_{\bxi,\bxi^\prime} 
- 
\frac{\langle k\rangle }{2\beta n} \label{eq:free_e} 
\end{eqnarray} 
This  involves the sublattice averages $\bra f(\bxi) \ket_{\bxi} = 
\sum_{\bxi} p(\bxi) f(\bxi)$, with $p_{\bxi} =\lim_{N\to\infty} 
|I_{\bxi}|/N$. Varying (\ref{eq:free_e}) with respect to $\{ 
P_{\bxi}(\bsigma), \hat{P}_{\bxi}(\bsigma)\}$ leads to the saddle 
point equations 
\begin{eqnarray} 
\widehat{P}_{\bxi}(\bsigma)&=&\sum_{\bsigma^\prime}\bra 
P_{\bxi^\prime}(\bsigma^\prime)e^{\frac{\beta}{\langle 
k\rangle}\phi(\bxi\cdot\bxi^\prime)(\bsigma\cdot\bsigma^\prime)}\ket_{\bxi^\prime}\label{eq:sp1}\\ 
P_{\bxi}(\bsigma)&=&\sum_{k}\frac{kp_k}{\langle 
k\rangle}\frac{e^{\beta  {\bf h}\cdot\bxi 
\sum_{\alpha=1}^n\sigma_\alpha}\widehat{P}^{k-1}_{\bxi}(\bsigma)}{\sum_{\bsigma}e^{\beta 
{\bf h}\cdot\bxi 
\sum_{\alpha=1}^n\sigma_\alpha}\widehat{P}^{k}_{\bxi}(\bsigma)}\label{eq:sp2} 
\end{eqnarray} 
Upon adopting the usual replica symmetric (RS) ans\"atze 
\cite{KS87,MP87,Mo98}, i.e. 
\begin{eqnarray} 
P_{\bxi}(\bsigma)&=&\int\! dh~ W_{\bxi}(h)\frac{e^{\beta 
h\sum_{\alpha=1}^n\sigma_\alpha}}{\big[2\cosh(\beta h)\big]^n} 
\label{eq:RS1} 
\\ 
\widehat{P}_{\bxi}(\bsigma)&=&\int\! du~ Q_{\bxi}(u)\frac{e^{\beta 
u\sum_{\alpha=1}^n\sigma_\alpha}}{\big[2\cosh(\beta u)\big]^n} 
\label{eq:RS2} 
\end{eqnarray} 
one finds  the saddle-point equations (\ref{eq:sp1}) and 
(\ref{eq:sp2}) reducing to 
\begin{eqnarray} 
\hspace{-1cm}Q_{\bxi}(u)&=&\bra \int\! dh~ 
W_{\bxi^\prime}(h)~\delta\left[u-\frac{1}{\beta}\tanh^{-1}\big[\tanh(\beta 
h)\tanh(\frac{\beta\phi(\bxi\cdot\bxi^\prime)}{\langle 
k\rangle})\big] \right]\ket_{\bxi^\prime} \label{eq:Q_RS} 
\\ \hspace{-1cm} 
W_{\bxi}(h)&=&\sum_{k}\frac{k p_k }{\langle k\rangle}\int 
\Big[\prod_{\ell=1}^{k-1}du_\ell~ 
Q_{\bxi}(u_\ell)\Big]\delta\left[h-\sum_{\ell=1}^{k-1} 
u_\ell-\sum_{\mu=1}^p h^\mu\xi^\mu\right] \label{eq:W_RS} 
\end{eqnarray} 
where $W_{\bxi}(h)$ and $Q_{\bxi}(u)$ are probability 
distributions for the effective cavity fields and propagated 
fields (messages) in sublattice $\bxi$, respectively 
\cite{LeVaVeZe02}. Insertion of (\ref{eq:RS1},\ref{eq:RS2}) into 
(\ref{eq:free_e}) gives us the  RS free energy per spin: 
\begin{eqnarray} 
&&\hspace{-2.0cm}\beta f=\langle k\rangle \bra \int\!  dhdu~ 
W_{\bxi}(h) Q_{\bxi}(u) \log\big[1+\tanh(\beta u)\tanh(\beta h) 
\big]\ket_{\bxi}\nonumber\\ &&\hspace{-1.3cm}-\frac{\langle 
k\rangle }{2}\bra\bra \int\!  dh dh^\prime 
W_{\bxi}(h)W_{\bxi^\prime}(h^\prime)\log\Big[1+\tanh(\beta 
h)\tanh(\beta 
h^\prime)\tanh\big[\frac{\beta\phi(\bxi\cdot\bxi^\prime)}{\langle 
k\rangle}\big]\Big]\ket\ket_{\bxi,\bxi^\prime}\nonumber\\ 
&&\hspace{-1.3cm}-\sum_{k}p_k\bra\int\Big[\prod_{\ell=1}^k du_\ell 
Q_{\bxi}(u_\ell)\Big]\log\Bigg(\frac{2\cosh\big(\beta 
\sum_{\ell=1}^ku_\ell+\beta\sum_{\mu=1}^p 
h^\mu\xi^\mu\big)}{\prod_{\ell=1}^k 2\cosh(\beta 
u_\ell)}\Bigg)\ket_{\bxi}\nonumber\\ 
&&\hspace{-1.3cm}-\frac{\langle k\rangle }{2}\bra\bra 
\log\cosh\big[\frac{\beta\phi(\bxi\cdot\bxi^\prime)}{\langle 
k\rangle}\big]\ket\ket_{\bxi,\bxi^\prime}-\langle k\rangle\log2 
\end{eqnarray} 
We may finally use the generating fields $h_\mu$ to find explicit 
expressions for the disorder-averaged pattern  recall overlaps 
$m^\mu=\lim_{N\to\infty}N^{-1}\sum_i \xi_i^\mu \odis \bra\sigma_i 
\ket \cdis_{\bc}=-(\partial f/\partial h^\mu)|_{{\bf h}=0}$: 
\begin{eqnarray} 
m^\mu=\sum_{k}p_k~\bra\xi^\mu\int\!\Big[\prod_{\ell=1}^k du_\ell 
Q_{\bxi}(u_\ell)\Big]\tanh\Big(\beta 
\sum_{\ell=1}^ku_\ell\Big)\ket_{\bxi} \label{mag} 
\end{eqnarray} 
In a similar manner one may derive an expression for the 
disorder-averaged RS spin-glass order parameter 
$q=\lim_{N\to\infty}N^{-1}\sum_i \odis \bra\sigma_i \ket^2 
\cdis_{\bc}$, upon adding a term of the form $\lambda 
\sum_{\alpha<\beta}\sigma_{\alpha}\sigma_{\beta}$ to the 
replicated Hamiltonian in (\ref{eq:Zn}). The result is 
\begin{equation} 
q=\sum_{k}p_k~\bra\int\!\Big[\prod_{\ell=1}^k du_\ell 
Q_{\bxi}(u_l)\Big]\tanh^2\Big(\beta 
\sum_{\ell=1}^ku_\ell\Big)\ket_{\bxi} \label{spglop} 
\end{equation} 
These expressions have a transparent interpretation, given that 
within the cavity formalism the local magnetization at a site $i$ 
is indeed given by $m_i=\tanh\big(\beta 
\sum_{\ell=1}^{k_{i}}u_\ell\big)$. 
 
\section{Phase diagram and order parameters} 
\label{phase} 
 
The paramagnetic (P) phase, where (\ref{mag}) and (\ref{spglop}) 
are zero, has $W_{\bxi}(x) = Q_{\bxi}(x) = \delta(x)$ for all 
$\bxi$. The recall phase (R) is defined by $m^\mu\neq 0$ for some 
$\mu$. In the spin-glass (SG) phase, $q>0$ but $m^\mu=0$ for all 
$\mu$. Transitions away from the P phase are expected to be second 
order, allowing us to find the P$\to$(R,SG) transitions via a 
simple continuous bifurcation analysis. In contrast, locating the 
SG$\to$R transition requires knowledge of the (nontrivial) 
functions $W_{\bxi}(h)$ and $Q_{\bxi}(u)$ in the R or SG regimes; 
to find these transitions we will solve 
(\ref{eq:Q_RS},\ref{eq:W_RS}) numerically 
 using a population dynamics algorithm.

Following \cite{WeCo03,PeSk03} we apply a bifurcation analysis to 
compute the second order transition lines away from the 
paramagnetic phase. By assuming the effective fields to be small, 
i.e. $\int\! dh~ W_\bxi(h) h^\ell={\cal O}(\epsilon^\ell)$  for 
all $\bxi$, and expanding equation (\ref{eq:Q_RS},\ref{eq:W_RS}) 
up to order $\epsilon^2$, one finds the following bifurcation 
conditions for transitions away from P: 
\begin{eqnarray} 
\hspace*{-15mm} {\rm P}\to{\rm R}: &~~~&  \int\! dh~ 
W_{\bxi}(h)h=\frac{\langle k^2 \rangle-\langle k \rangle} {\langle 
k \rangle}\bra\tanh[\frac{\beta\phi(\bxi\cdot\bxi^\prime)}{\langle 
k \rangle}] \int\! dh~ W_{\bxi^\prime}(h)h \ket_{\bxi^\prime} 
\\ 
\hspace*{-15mm} {\rm P}\to{\rm SG}: &~~~& \int\! dh~ 
W_{\bxi}(h)h^2= \frac{\langle k^2 \rangle-\langle k 
\rangle}{\langle k 
\rangle}\bra\tanh^2[\frac{\beta\phi(\bxi\cdot\bxi^\prime)}{\langle 
k \rangle}]  \int\!  dh~ W_{\bxi^\prime}(h)h^2 \ket_{\bxi^\prime} 
\end{eqnarray} 
For  $p_\bxi=2^{-p}$ (random patterns) one can find the 
 eigenvalues of 
the relevant matrices 
\begin{eqnarray} 
M_{\bxi\bxi^\prime}&=&2^{-p}~\frac{\langle k^2 \rangle-\langle k 
\rangle}{\langle k 
\rangle}\tanh\Big[\frac{\beta\phi(\bxi\cdot\bxi^\prime)}{\langle k 
\rangle}\Big]\\ Q_{\bxi\bxi^\prime}&=&2^{-p}~\frac{\langle k^2 
\rangle-\langle k \rangle}{\langle k 
\rangle}\tanh^2\Big[\frac{\beta\phi(\bxi\cdot\bxi^\prime)}{\langle 
k \rangle}\Big] 
\end{eqnarray} 
and the conditions for a second order transition become, see 
\cite{WeCo03}: 
\begin{eqnarray} 
\hspace*{-5mm} {\rm P}\to{\rm R}: &~~~& \frac{\langle k^2 
\rangle-\langle k \rangle}{\langle k 
\rangle}\frac{2^{-p}}{p}\sum_{n=0}^{p}{p \choose 
n}(p-2n)\tanh\Big[\frac{\beta\phi(p-2n)}{\langle k \rangle}\Big]=1 
\label{eq:bifr} \\ \hspace*{-5mm} 
 {\rm P}\to{\rm SG}: &~~~& 
\frac{\langle k^2 \rangle-\langle k \rangle}{\langle k 
\rangle}2^{-p}\sum_{n=0}^{p}{p \choose 
n}\tanh^2\Big[\frac{\beta\phi(p-2n)}{\langle k \rangle}\Big]=1 
\label{eq:bifsg} 
\end{eqnarray} 
 For a Poissonian degree distribution one has $\bra k^2\ket=\bra k\ket^2+\bra k\ket$, and we recover the results in 
\cite{WeCo03}. For non-Poissonian distributions, equation 
(\ref{eq:bifr}) predicts an enlargement of the retrieval phase 
when the degree distribution has fat tails, e.g. for power-law 
distributions $p(k)\sim k^{-\gamma}$. As noted by 
\cite{LeVaVeZe02}, if the second moment of the degree distribution 
is not finite (e.g. for power-law distributions with $\gamma \leq 
3$), there is always a retrieval phase in the thermodynamic limit 
(unless one re-scales temperature first). For a single pattern, 
the retrieval phase boundary is given, in accordance with 
\cite{LeVaVeZe02}, by 
\begin{equation} 
\beta_c=-\frac{\langle k\rangle}{2}\log\Big(1-\frac{2\langle k \rangle}{\langle k^2 \rangle}\Big) 
\end{equation} 
(our expression differs slightly, due to our rescaling of the 
bonds by a factor $\langle k \rangle^{-1}$).\vsp

To find our order parameters we solve equations 
(\ref{eq:Q_RS},\ref{eq:W_RS}) 
 numerically, using a population dynamics 
algorithm \cite{PM00,BergSellito}. We have two distributions for 
each of our $2^p$ sublattices, so the required CPU time grows 
exponentially with $p$. 
 However, one may exploit sublattice symmetries, 
 especially when the system is condensed in a single pattern 
retrieval state \cite{progress}. If the first pattern is 
condensed, numerical and analytical evidence (as yet short of a 
proof) suggests the solution of (\ref{eq:Q_RS},\ref{eq:W_RS}) to 
have $W_{\bxi}(h)=W_{\xi^1}(h)$. Moreover, for random patterns one 
expects the symmetry $W_{\xi^1}(h)=W{(\xi^1}h)$. Insertion of this 
ansatz (and a similar one for $Q_{\bxi}$) into 
(\ref{eq:Q_RS},\ref{eq:W_RS}) leads us to 
\begin{eqnarray} 
W(h)&=&\sum_{k}\frac{k p_k }{\langle k\rangle}\int\! 
\Big[\prod_{\ell=1}^{k-1}du_\ell 
~Q(u_\ell)\Big]\delta\Big[h-\sum_{\ell=1}^{k-1} u_\ell\Big] 
\label{eq:condwh} 
\\ 
Q(u)&=&\frac{1}{2^{p-1}}\sum_{n=0}^{p-1}{p-1 \choose n}\int\! dh~ 
W(h)\nonumber\\ 
&&\times~\delta\Big[u-\frac{1}{\beta}\tanh^{-1}\big[\tanh(\beta 
h)\tanh(\frac{\beta\phi(p-2n)}{\langle k\rangle})\big] \Big] 
\label{eq:condqu} 
\end{eqnarray} 
Inserting (\ref{eq:condqu}) into (\ref{eq:condwh}) then gives the 
relatively simple expression 
\begin{eqnarray} 
W(h)&=&\sum_{k}\frac{k p_k }{\langle k\rangle} 
\prod_{\ell=1}^{k-1} 
\Bigg[\frac{1}{2^{p-1}}\sum_{n_\ell=0}^{p-1}{p-1 \choose n_l} 
\int\! dh_\ell~ W(h_\ell)\Bigg]\nonumber\\ &&\times ~ 
\delta\Bigg\{h-\frac{1}{\beta}\sum_{\ell=1}^{k-1}\tanh^{-1}\Big[\tanh(\beta 
h_\ell)\tanh[\frac{\beta\phi(p-2n_\ell)}{\langle k\rangle}]\Big] 
\Bigg\} 
\end{eqnarray} 
 
\section{Comparison with simulations} 
 
\begin{figure}[t] 
\setlength{\unitlength}{0.6mm} \vspace*{3mm} \hspace*{10mm} 
\begin{picture}(250,100) 
\put(0,-10){\includegraphics[height=100\unitlength, 
width=110\unitlength] {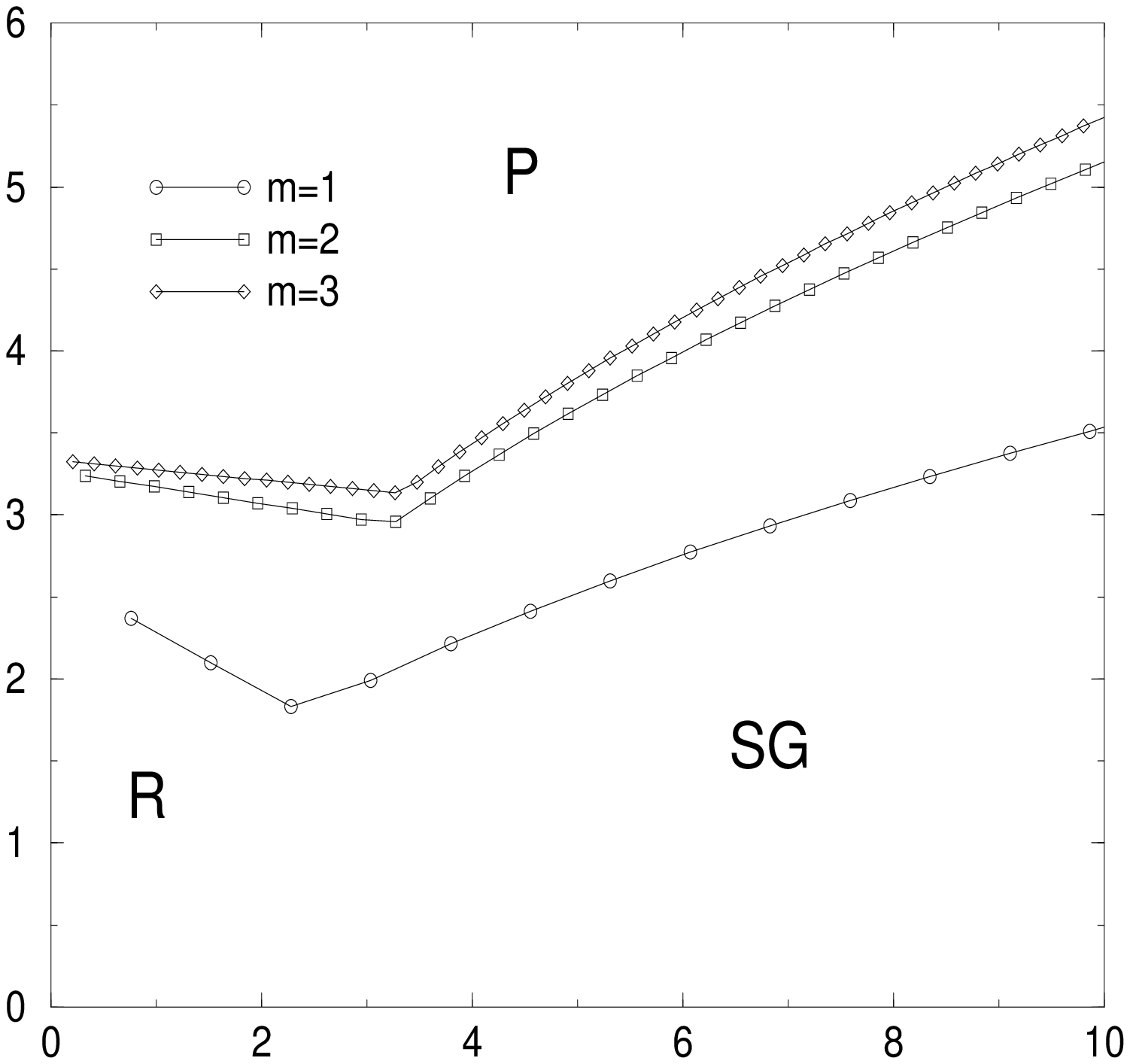}} 
\put(125,-10){\includegraphics[height=100\unitlength, 
width=110\unitlength] {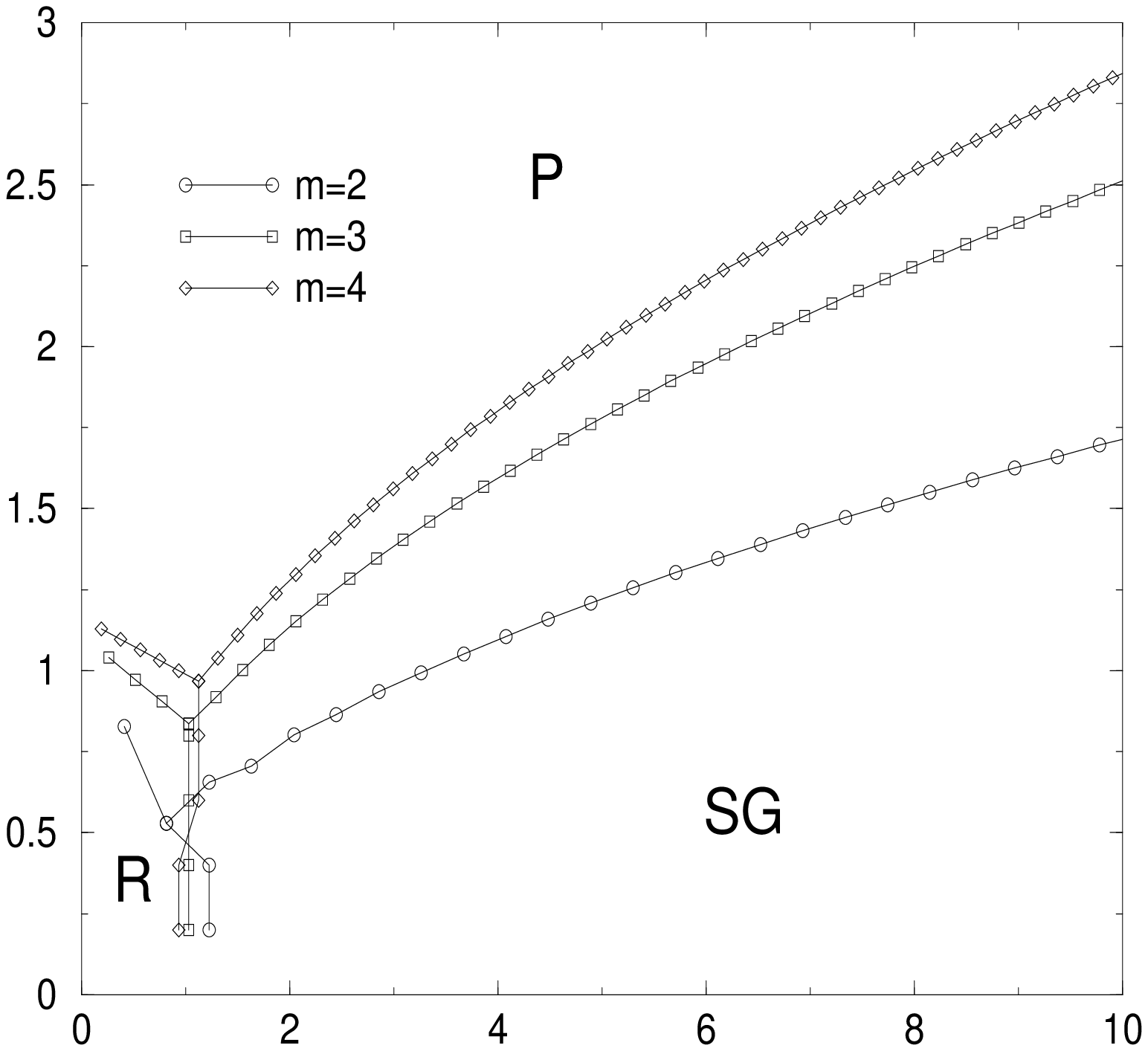}} 
 \put(56,-9){\small $\alpha$} \put(183,-9){\small $\alpha$} \put(-8,52){\small $T$} 
\put(120,52){\small $T$} 
\end{picture} 
\vspace*{0mm} \caption{RS phase diagrams for $p(k) \sim 
k^{-\gamma}$ with $p(k)=0$ for $k<m$, where $\gamma=3.1$ (left) 
and $\gamma=4$ (right). Phase boundaries found by bifurcation 
analysis, separating paramagnetic (P) from retrieval (R) and spin 
glass (SG) phases. For $\gamma=4$ the R$\to$SG phase boundary is 
calculated for $T$ values which are multiples of $0.2$, and is 
found via population dynamics (with a condensed ansatz). For 
$\gamma=3.1$, average connectivity values corresponding to $m=1$ 
(circles), $m=2$ (squares) and $m=3$ (diamonds) are $\langle k 
\rangle = 1.318$, $\langle k \rangle = 3.055$ and $\langle k 
\rangle = 4.898$ respectively. For $\gamma = 4$, average 
connectivity values corresponding to $m=2$ (circles), $m=3$ 
(squares) and $m=4$ (diamonds) are $\langle k \rangle = 2.454$, 
$\langle k \rangle = 3.887$ and $\langle k \rangle = 5.352$ 
respectively.} \label{phasefig} 
\end{figure} 
 
In figure \ref{phasefig} we present the resulting RS phase 
diagrams in the $(\alpha, T)$-plane, where $\alpha = p/\langle 
k\rangle$ for Hopfield-type networks with power-law degree 
distribution $p(k) \sim k^{-\gamma}$. The distributions $p(k)$ are 
characterized by $\gamma=3.1$ and $\gamma=4$ respectively, and by 
a variable $m$ which defines a lower cutoff ($p(k)=0$ for $k<m$). 
At high $T$ one finds the paramagnetic (P) phase. At sufficiently 
low $T$ one finds a retrieval (R) phase (small $\alpha$) or  a 
spin glass (SG) phase (large $\alpha$). The value $\gamma=3.1$ is 
close to the critical value $\gamma_c=3$ below which there is no 
paramagnetic phase, yet here the phase diagram is found to be 
similar to that corresponding to Poissonian graphs \cite{WeCo03} 
(with re-scaled  values of $T$ and $\alpha$). In the $\gamma=4$ 
phase diagram we also indicate the location of the R$\to$SG 
transition, resulting from a population dynamics calculation. The 
P$\to$R and P$\to$SG  boundaries were found by solving the 
bifurcation equations (\ref{eq:bifr},\ref{eq:bifsg}) numerically. 
For small values of the order parameters, i.e. large $T$ or 
$\alpha$, finding accurate numerical values for the order 
parameters becomes increasingly difficult. Hence, for large 
fluctuations in the connectivity (i.e. large values of $\langle 
k^2 \rangle$), one cannot expect accurate results for the R$\to$SG 
transition on the basis  of a population dynamics algorithm. This 
is why we have omitted the R$\to$SG line in the $\gamma = 3.1$ 
phase diagram. 
 
\begin{figure}[t] 
\vspace{17mm} \hspace{10mm} 
\begin{picture}(1050, 100) 
\setlength{\unitlength}{0.6mm} \put(-2,-30){ 
\includegraphics[height=101\unitlength, width=110\unitlength] 
{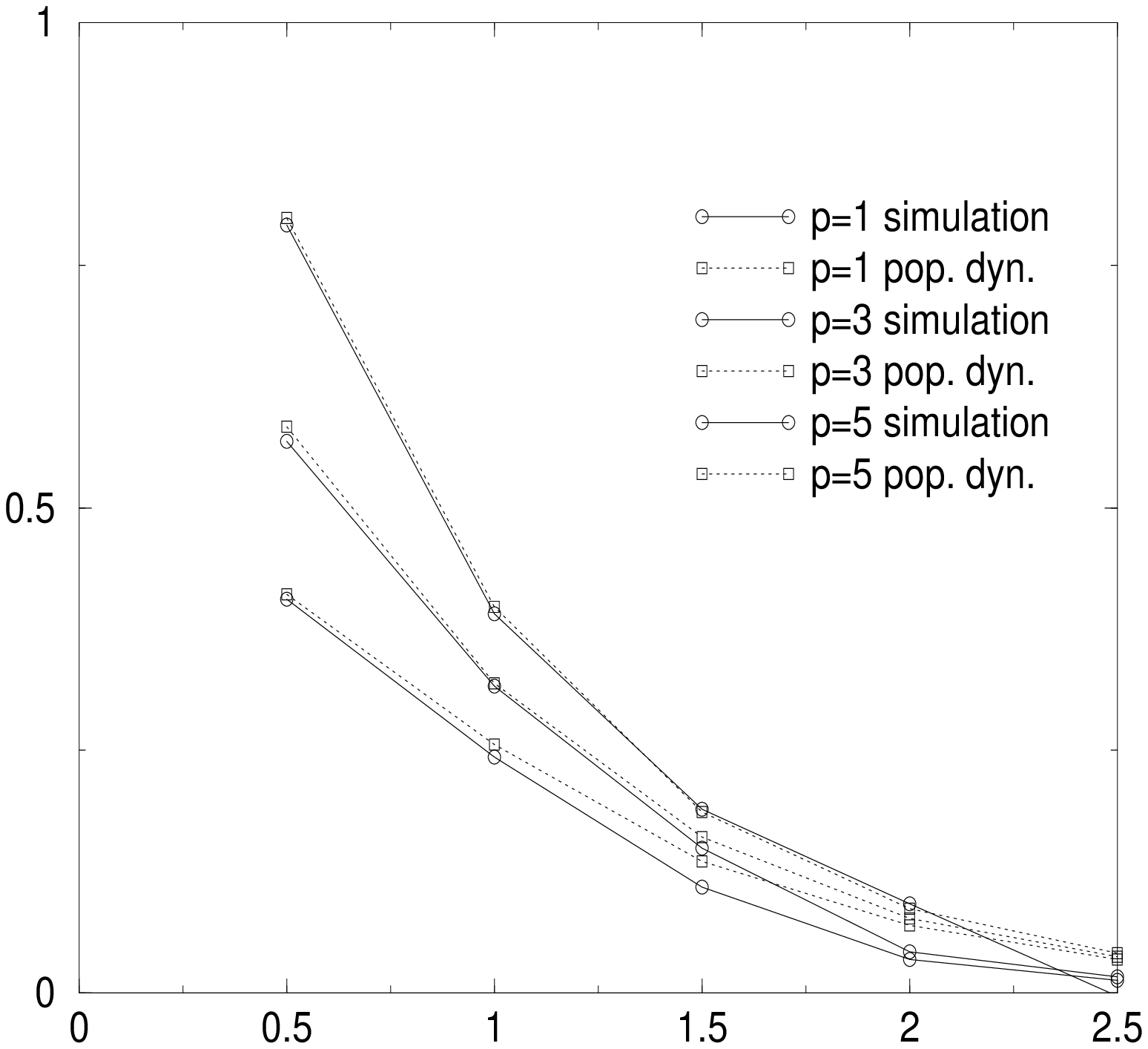}} \put(125,-30){ 
\includegraphics[height=101\unitlength, width=110\unitlength] 
{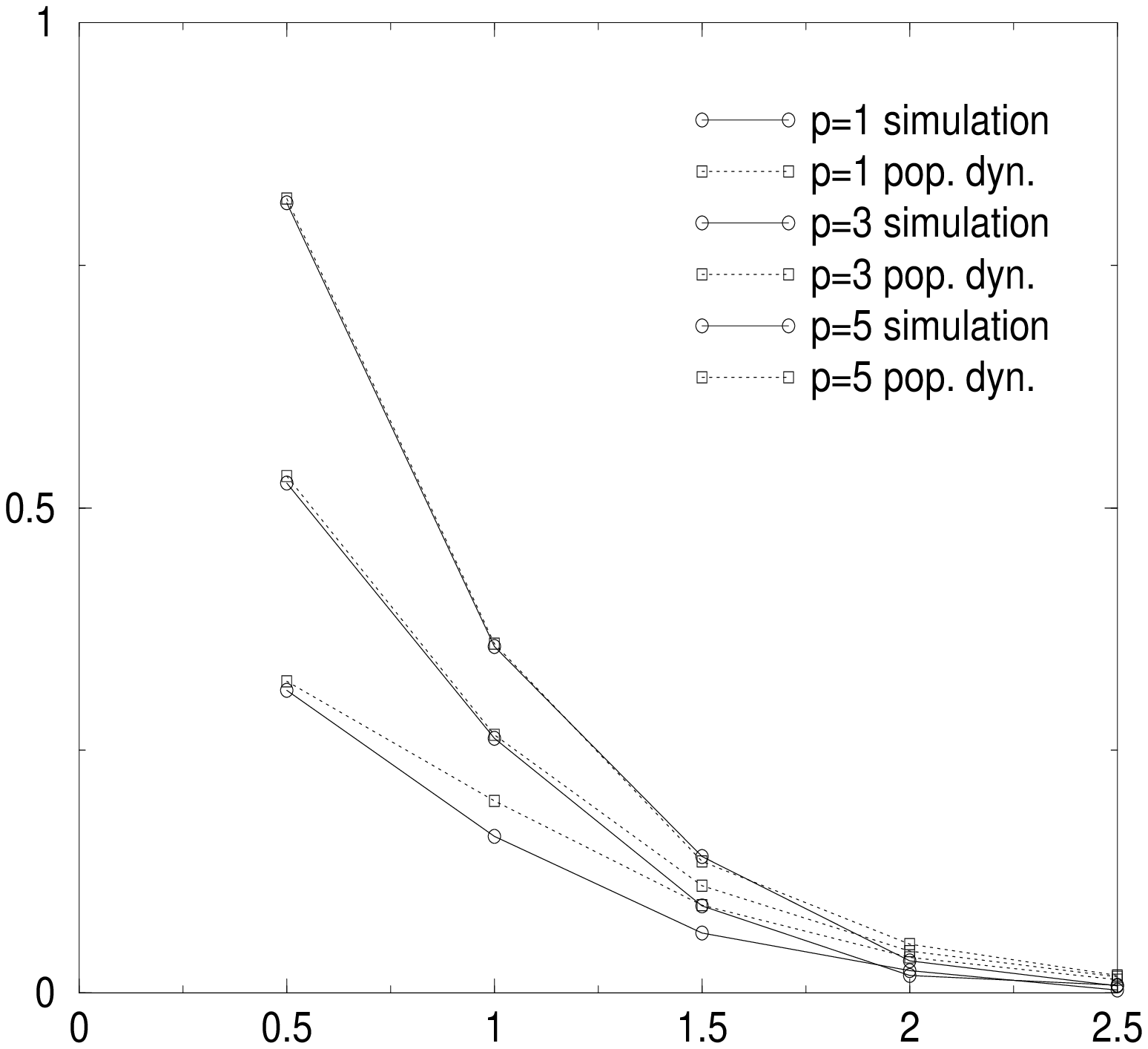}} 
 
\put(-8,32){\small $m^1$} \put(121,32){\small $m^1$} 
\put(54,-30){\small $T$} \put(183,-30){\small $T$} 
\end{picture} 
\vspace*{13mm} \caption{Condensed recall overlap $m^1$ as 
calculated from our RS theory via population dynamics (squares) 
versus numerical simulations (circles), for a graph with $p(k) 
\sim [k(k+1)(k+2)]^{-1}$ (left), and a graph with $p(k) \sim 
k^{-3}$ (right), both with $p(k)=0$ for $k<3$. The corresponding 
average connectivities are $\langle k \rangle =6$ (left) 
and $\langle k \rangle = 5.123$ (right). All data shown are 
averages over $10$ runs,  in both simulations ($N=10^4$ spins) and 
in population dynamics (with populations of size $N=10^4$).} 
\label{mTsimpop} 
\end{figure} 
 
\begin{figure}[t] 
\hspace*{18mm}\vspace*{5mm} 
\begin{picture}(270,170) 
\put(85,0){\epsfig{file=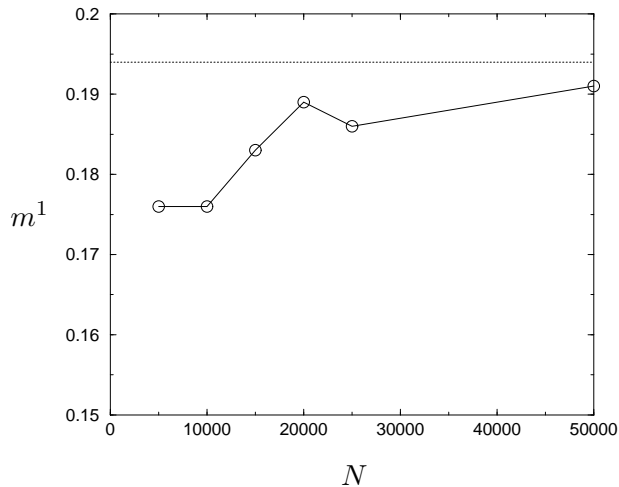, width=210\unitlength}} 
\put(190,-18){\small $N$} \put(65,80){\small $m^1$} 
\end{picture} 
\vspace{3mm} \caption{Average retrieval overlap $m^1$ over 10 
simulations as a function of system size $N$, for connectivity 
degree distribution $p(k) \sim k^{-3}$ (with $p(k)=0$ for $k<3$, 
at $T=1$ and $p=5$).  The dotted line corresponds to the value 
predicted by our RS population dynamics.} \label{finsiz} 
\end{figure}

In figure \ref{mTsimpop} we show as function of temperature the 
recall overlap $m^1$ (for states where only pattern 1 is 
condensed) as obtained from a population dynamics calculation, 
together with the measurements of $m^1$ in numerical simulations. 
Both simulation and population dynamics had $N=10^4$ and all data 
are averages over 10 runs.  The left figure refers to  $p(k) \sim 
[k(k+1)(k+2)]^{-1}$, corresponding to the degree distribution 
resulting from a Barabasi-Albert algorithm for network growth 
\cite{BarAl,BarAlrev}. It should be noted, however, that the 
degree-degree correlation generated by the latter algorithm 
differs from the one  (\ref{eq:degdegcor}) in our present model 
 (see e.g. \cite{GrSnMi04}). Consequently, a different algorithm, 
similar to the one described in \cite{FarDer}, had to be used 
here. First, a set $\{ k_1, \ldots, k_N\}$ is generated in 
accordance with the distribution $p(k)$. One then chooses randomly 
two sites $i$ and $j$ with probabilities $p(i) = 
\frac{k_i}{\sum_{j} k_j} \sim p(k_i) k_i$. Unless these two sites 
coincide or already share a bond, they are connected. If site $i$ 
already has $k_i$ connections, it is excluded from the process to 
speed up the algorithm. This process is repeated until all 
connectivities have the correct value. We also used this algorithm 
to generate  power law distributed graphs with $\gamma=3$ (right 
figure). For both architectures, the overlaps $m^1$ have been 
plotted for networks with $p=1$, $p=3$ and $p=5$. At low 
temperature, the results of the simulation agree well with the 
population dynamics results. For the values of $p$ used in figure 
(\ref{mTsimpop}) one has pattern retrieval at sufficiently low 
temperatures. In fact our theory claims that for $p(k)\sim 
k^{-\gamma}$ with $\gamma \leq 3$ one will have retrieval at any 
$T$. For large values of $p$ and $T$, however, the overlaps, 
although indeed nonzero, become smaller and hence our numerical 
accuracy decreases. Moreover, the equilibration times in the 
simulations grow rapidly as $p$ increases. The discrepancies at 
high temperatures and large $p$ are, we believe, due to finite 
size effects.  In figure \ref{finsiz} we show 
 that the agreement between theory and simulations indeed improves for larger system sizes. 
 
\vspace{-2mm} 
 
\section{Conclusions} 
 
\label{conclusion} 
 
We have solved  attractor neural network models on random graphs 
with arbitrary connectivity distributions $p(k)$, using the 
replica method within RS ansatz, in the spirit of 
\cite{WoSh87,LeVaVeZe02,WeCo03}. The RS order parameters are the 
effective cavity field distributions $W_{\bxi}(h)$ in each 
sublattice, or equivalently, the distributions of messages 
$Q_\bxi(u)$. Second order phase transitions from the paramagnetic 
(P) phase to  a retrieval (R) or spin glass (SG) phase could be 
derived explicitly, given the assumption that these transitions 
are second order and  provided the second moment $\langle k^2 
\rangle$ of the connectivity degree distribution of the graph is 
finite. The overlap and spin glass order parameters in each phase 
can in principle be calculated via a population dynamics 
algorithm. The latter is limited by numerical accuracy when the 
values of these order parameters are small (as for  large $T$ and 
$\alpha$). We find that the retrieval region in the phase diagram 
is larger for fat-tailed degree distributions than for those with 
exponential decay (e.g. Poisonian), but it is not clear whether 
this can be exploited in associative memories since it goes at the 
cost of the magnitude of the retrieval overlaps. The possible 
occurrence of replica symmetry breaking is beyond the scope of 
this paper. Within our numerical accuracy, we can conclude that 
upon comparing the results of our  replica symmetric theory 
(including population dynamics) to numerical simulations, for 
degree distributions $p(k)\sim [k(k+1)(k+2)]^{-1}$ and $p(k) \sim 
k^{-\gamma}$, we obtain satisfactory agreement. 
 
\section*{Acknowledgment} 
 
This study was initiated during an informal Finite Connectivity 
Workshop at King's College London in November 2003. TN, IPC, NS 
and BW acknowledge financial support from the State Scholarships 
Foundation (Greece), the Fund for Scientific Research (Flanders, 
Belgium), the ESF SPHINX programme and the Ministerio de
Educaci\'on, Cultura y Deporte (Spain, grant SB2002-0107), 
and the FOM Foundation (Fundamenteel Onderzoek der Materie, 
The Netherlands), respectively.

\end{document}